\documentclass[preprint,12pt]{elsarticle}

\usepackage{amssymb}
\usepackage{amsmath}

\usepackage{lineno}

\journal{Nuclear Instruments and Methods in Physical Research A}

\begin{document}

\begin{frontmatter}

\title{
Nuclotron internal target polarimeter for the measurements of the deuteron and proton beam polarization}

\author[1]{I.~S.~Volkov}\ead{isvolkov@jinr.ru}  
\author[1]{V.~P.~Ladygin}
\author[1]{A.~V.~Averyanov}
\author[1]{E.~V.~Chernykh}
\author[2]{D.~Enache}
\author[1]{Yu.~V.~Gurchin}
\author[1]{A.~Yu.~Isupov}
\author[3]{M.~Janek}
\author[1,2]{J.-T.~Karachuk}
\author[1]{D.~O.~Krivenkov}
\author[1]{P.~K.~Kurilkin}
\author[1]{A.~N.~Livanov}
\author[1]{S.~M.~Piyadin}
\author[1]{S.~G.~Reznikov}
\author[1]{Ya.~T.~Skhomenko}
\author[1]{A.~A.~Terekhin}
\author[1]{A.~V.~Tishevsky}
\author[4]{T.~Uesaka}
\author[5]{I.~E.~Vnukov}
\author[]{(DSS-Collaboration)}

\affiliation[1]{organization={Joint Institute for Nuclear Research},
            addressline={6 Joliot-Curie}, 
            city={Dubna},
            postcode={141980}, 
            state={Moscow region},
            country={Russia}}
\affiliation[2]{organization={National Institute for R\&D in Electrical Engineering ICPE-CA},
            addressline={313 Splaiul Unirii, District 3}, 
            city={Bucharest},
            postcode={030138},
            country={Romania}}
\affiliation[3]{organization={Physics Department, University of \v{Z}ilina},
            addressline={Univerzitn\'{a} 8215/1}, 
            city={\v{Z}ilina},
            postcode={010 26},
            country={Slovakia}}
\affiliation[4]{organization={Nishina Center for Accelerator-Based Science, RIKEN},
            city={Wako},
            postcode={308015},
            country={Japan}}
\affiliation[5]{organization={Belgorod State National Research University},
            addressline={85 Pobedy St.}, 
            city={Belgorod},
            postcode={308015},
            country={Russia}}

\begin{abstract}
Studies of  spin-dependent effects at the Nuclotron/NICA accelerator complex at JINR require 
precise measurements of the deuteron and proton beam polarization.
The vector polarization of the deuteron beam was measured at 
the energies of 200, 500, 550, and 650 MeV/nucleon by a
detection system of scintillation counters placed at the Nuclotron internal target. 
Considering the deuteron beam as a beam of weakly bound protons and
neutrons, the asymmetries of scattering of protons from 
deuterons on polyethylene and carbon targets were determined. 
The polarization of the polarized proton beam accelerated for the first time at 
the Nuclotron up to 500 MeV was also
measured. 
\end{abstract}


\end{frontmatter}

\section{Introduction} 
\label{sec:introduction}

The investigation of the nuclear matter spin structure remains a crucial aspect of contemporary scientific research. The major directions of spin studies  at the Laboratory of High Energy Physics (LHEP) at JINR are the spin structure of two-nucleon and three-nucleon short-range correlations~\cite{dss1,dss2,dss3} and  applications for high energy focal plane polarimetry to measure the nucleon electromagnetic formfactors at large $Q^2$~\cite{basilev_2020}  
using internal and extracted polarized beams, respectively.  
The commissioning of the Spin Physics Detector (SPD)~\cite{SPD_TDR,SPD2025} at the Nuclotron-based Ion Collider fAcility (NICA)~\cite{nica} will significantly expand the scope of spin studies at JINR. The main objectives of the  
SPD scientific program are  the gluon spin structure of the proton and deuteron~\cite{spd_gluon} and 
other spin effects and polarization phenomena~\cite{spd_stage1} using polarized proton and deuteron beams at the luminosity up to 10$^{32}$~cm$^{-2}\cdot$s$^{-1}$ and at the collision energy up to 27~GeV. 
The reliable measurement of spin effects at the Nuclotron/NICA complex inherently depends on a good knowledge and  continuous monitoring of the beam polarization throughout the data taking.

For these purposes, proton and deuteron beam polarimeters 
measuring the asymmetry of  
quasi-elastic $pp$-scattering on hydrogen in  
polyethylene (CH$_2$) and carbon targets are often used. 
This classical method is based on the fact that
the analyzing powers of elastic and quasi-elastic $pp$-scattering show no   significant  difference within the experimental precision achieved over a broad energy range ~\cite{Ball1987,Ball_1999_pp_el_quasiel}. 
This allows one to measure both the proton beam polarization and the deuteron beam vector polarization, as the detectors are arranged according to $pp$-elastic scattering kinematics.
Such polarimeters were used previously at ANL ~\cite{spinka}, KEK ~\cite{kek}, SATURNE-II ~\cite{saturn-2}, COSY ~\cite{Altmeier2005}, and the JINR Synchrophasotron \cite{azhgirey2003}.  Nowadays the upgraded version of the 
polarimeter \cite{azhgirey2003} is currently used  
to measure the vector polarization of the extracted deuteron beam at the Nuclotron 
\cite{azhgirey2005,basilev_2020}.

The paper presents the polarimeter based on  asymmetry measurements in  $pp$-quasi-elastic scattering at intermediate energies at the Nuclotron Internal Target Station (ITS)~\cite{ITS}.
The results of the  deuteron beam vector polarization  at  200, 500, 550, and 650 MeV/nucleon, as well as the proton beam polarization at  500 MeV, are presented.
The paper is organized as follows.
The experimental method  is described in Section~\ref{sec:experiment_scheme}.
A detailed description of the event selection with the carbon background subtraction procedure is provided in Section~\ref{sec:useful_events}. The procedure  for measuring the deuteron beam vector polarization  
using $pp$-quasi-elastic scattering asymmetries is outlined in Section~\ref{sec:dpol}.
Results on the effective analyzing power and polarimeter figure of merit are reported in 
Section~\ref{sec:FoM}. The polarization measurements of the polarized proton beam, accelerated at the 
Nuclotron for the first time, are described in  Section~\ref{sec:proton_beam_results}.
Conclusions are drawn in the last Section.

\section{The experimental method} 
\label{sec:experiment_scheme}

%
The beams of polarized deuterons and protons for the experiment were provided by the Source of Polarized Ions (SPI)~\cite{Fimushkin_2016_SPI, Belov_2017_SPI} developed for the Nuclotron/NICA accelerator complex.
SPI is an atomic beam-type polarized ion source with a charge-exchange plasma ionizer
and a storage cell in the ionization region.  Nuclear polarization is
provided via radio-frequency (RF) hyperfine structure transitions.
Three spin modes of   SPI were used for the deuteron beam: unpolarized, "2--6", and "3--5", with maximal theoretical values ($p_{Z}$,$p_{ZZ}$) = $(0,0)$, $(+1/3,+1)$, and $(+1/3,-1)$, respectively.
SPI provided only two spin modes for the proton beam: 
unpolarized and "1--3" with maximal theoretical values ($P$) = $(0)$, $(+1)$, respectively. 
The spin modes were changed cyclically and spill-by-spill. 
The polarizations ($p_{Z}$,$p_{ZZ}$) refer to the frame of the ion source, while the measured transverse component  perpendicular to beam-circulation plane of the Nuclotron is denoted as $p_y$ according to the Madison convention \cite{madison}. 

The typical beam intensity in the Nuclotron ring was $\sim$1--5$\cdot$10$^8$
deuterons per spill with a duration of $\sim$1~s regardless of  the
spin mode. The repetition rate was  1/6 Hz at all energies. 
The Nuclotron ITS ~\cite{ITS}  
consists of a spherical scattering chamber and
a target sweeping system. The scattering chamber is fixed on the
flanges of the Nuclotron beam pipe.
A disk with 6 targets made of various materials ($\text{CH}_2$, $\text{C}$, $\text{Ag}$, $\text{W}$, etc.)  is located on the axle of a stepper motor  inside the vacuum chamber.  
Targets are fixed between the rim and hub of the disk at  designated positions.
During beam acceleration, the disk is turned with an empty place, and upon reaching the required energy, the disk is rotated to bring the desired target onto the beam trajectory~\cite{ITS_DAQ}.
A polyethylene film with a thickness of 10~$\mu$m was used as a hydrogen target. 
The carbon target, consisting of 10 twisted filaments, each 8~$\mu$m thick, was used to evaluate the background from carbon content in the polyethylene.
The effect on   hydrogen was obtained using CH$_2$ --C subtraction. 
A signal from the target position monitor  (TPM) \cite{TPM} was used to tune
the accelerator parameters to bring the interaction point
close to the center of the ITS chamber.  The uncertainty of the interaction point 
is estimated to be   $\pm$1 mm. The details of the target position monitor operation are
described in Ref. \cite{TPM}.
 
\begin{figure}[h]
    \centering
    \includegraphics[width=100mm]{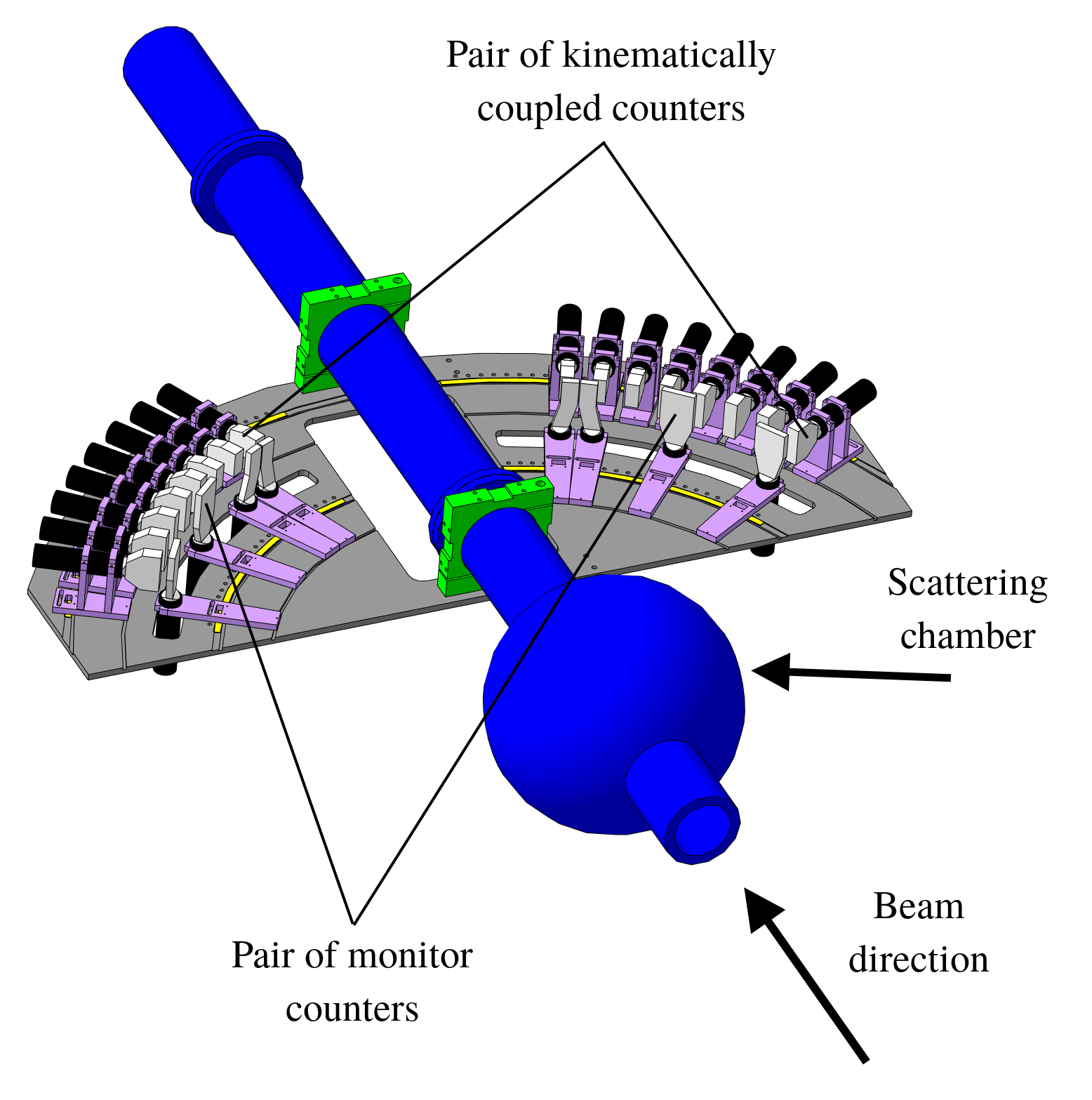}
    \caption{
        \label{fig:setup}
Layout of the $pp$- polarimeter setup installed downstream the  Nuclotron ITS \cite{ITS}
spherical chamber.  The plastic scintillation counters are placed according to $pp$-elastic scattering kinematics in the beam-circulation plane of the Nuclotron.
    }
\end{figure}

\begin{table}[h]
    \caption{
        \label{tab:detectors_prop}
        Plastic scintillation counters used for the Nuclotron ITS $pp$- polarimeter. 
        }
    \centering
    \begin{tabular}{ccccc}
        \hline
         c.m. angle  
         &Width 
         &Height  
         &Thickness  
         &Radius  \\
        ($^\circ$) & (mm) & (mm) & (mm) & (mm) \\
        \hline
        55, 65, 75, 85      & 20 & 40 & 20 & 580\\
        58, 62              & 24 & 40 & 10 & 555\\
        90                  & 50 & 60 & 10 & 555\\
        95, 105, 115, 125   & 20 & 60 & 20 & 580\\
        120                 & 50 & 40 & 10 & 555\\
        \hline
    \end{tabular}
\end{table}

A schematic view of the Nuclotron ITS $pp$-polarimeter setup designed for the
deuteron and proton beam polarization measurements in a wide energy range is
shown in Fig.~\ref{fig:setup}. Secondary particles from the interaction of the beam with the target were registered by  plastic scintillation counters.
The $pp$-polarimeter detection system is based on 
the partial use of the  scintillation counters from the 270 MeV deuteron beam polarimeter 
 described elsewhere ~\cite{dp270}. 
A detector support with 24 mounted plastic
scintillation counters according to $pp$-elastic scattering kinematics
at a given energy is placed downstream the ITS spherical
chamber  in the beam-circulation plane of the Nuclotron. 
Each plastic scintillation counter was coupled to a photo-multiplier tube Hamamatsu H7416MOD.
Sizes of the detectors, their distances from the target position, and their
setting angles in the center of mass  (c.m.) are listed in Table \ref{tab:detectors_prop}. 
Note that the angles in the c.m. are the same for 
all deuteron and proton beam energies.
One pair of detectors was placed to
register two protons from  $pp$-quasi-elastic scattering at 90$^\circ$
in the c.m.,  where the analyzing power $A_y$ value equals to zero. 
Therefore, this yield is independent of the beam polarization value and can be used
as a luminosity monitor of the beam-target collisions. 
 
The taking and recording of  the data from detectors were provided by the VME based data-
acquisition  (DAQ) system  \cite{DSS_DAQ}. Anode signals from the photo-multiplier tubes were fed
into the inputs of the TQDC-16   modules \cite{tqdc16}, each of which is a 16-channel discriminator, 
multi-hit time-stamping time-to-digit converter (TDC), and waveform digitizer.  
It is used to measure the signal arrival time, charge, and pulse shape.
Each TQDC-16 also has a  local trigger output which can be used in the trigger logic. 
The coincidences of any of
detectors placed on the left with any of the  detectors placed on the right from the 
beam direction were  used to trigger the data taking. 
The TTCM (Trigger, Timing and Control Module) module \cite{TTCM} and  the FVME controller \cite{FVME}
were used for clock, triggering, and data recording purposes during the experiment with the 
polarized deuteron beam. 
The upgraded DAQ system with the FVME2 controller \cite{FVME2},  trigger logic modules 
FVME2TMWR  \cite{FVME2TMWR}  and 
U40VE-TM \cite{U40VE-TM} were used to measure the proton beam polarization. 
All modules were mounted in the WIENER VME64x6021 crate. Data exchange was performed  using the optical transmission line.   
 
\section{Event selection} 
\label{sec:useful_events}

\begin{figure}[hbtp]
     \centering
     \includegraphics[width=120mm]{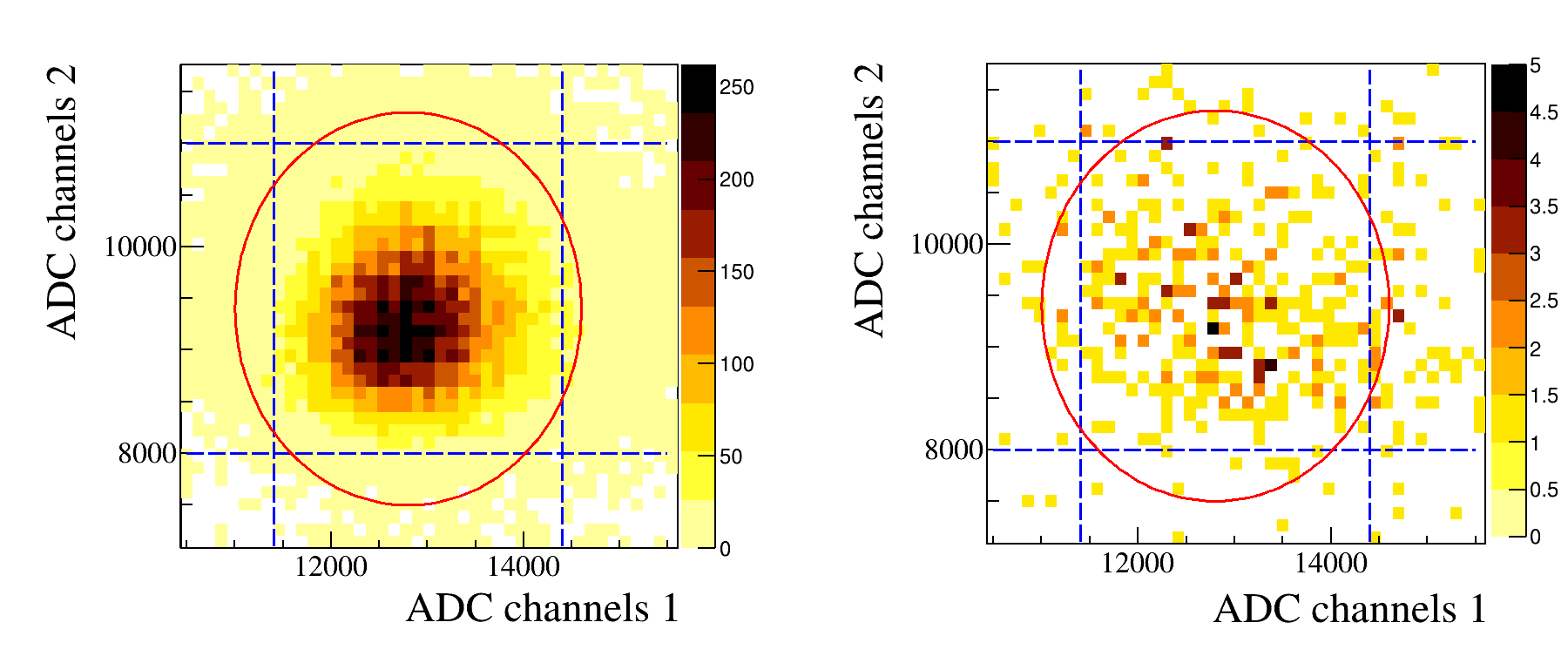}
     \caption{
Energy loss correlation for the counter pair at 75$^\circ$ in the c.m. at the energy of 550 MeV/nucleon.
The lines are explained in the text.
    } 
\label{fig:adc_cut}
\end{figure}

The $pp$-quasi-elastic events were selected by using the energy loss
correlation and time-of-flight difference for the signals from 
the scintillation counters for the scattered and recoil protons positioned according to   
$pp$-elastic scattering kinematics at a given energy (see Table~\ref{tab:detectors_prop}).
The TPM \cite{TPM} information was also used  to filter out the events originating far from the beam axis.
The energy loss correlations for two protons for the scattering angle of 75$^\circ$ in the c.m. and at 
the energy of 550 MeV/nucleon for $\text{CH}_2$ and carbon targets are shown in the left and right
panels of  Fig.~\ref{fig:adc_cut}, respectively. One can see a prominent
locus corresponding to the $pp$-quasi-elastic  events in the left panel. The solid line is a
graphical cut for the selection of    $pp$-quasi-elastic events. Dashed lines denote the selection criteria 
for the $\text{CH}_2-\text{C}$ subtraction procedure.
 
\begin{figure}[hbtp]
     \centering
     \includegraphics[width=120mm]{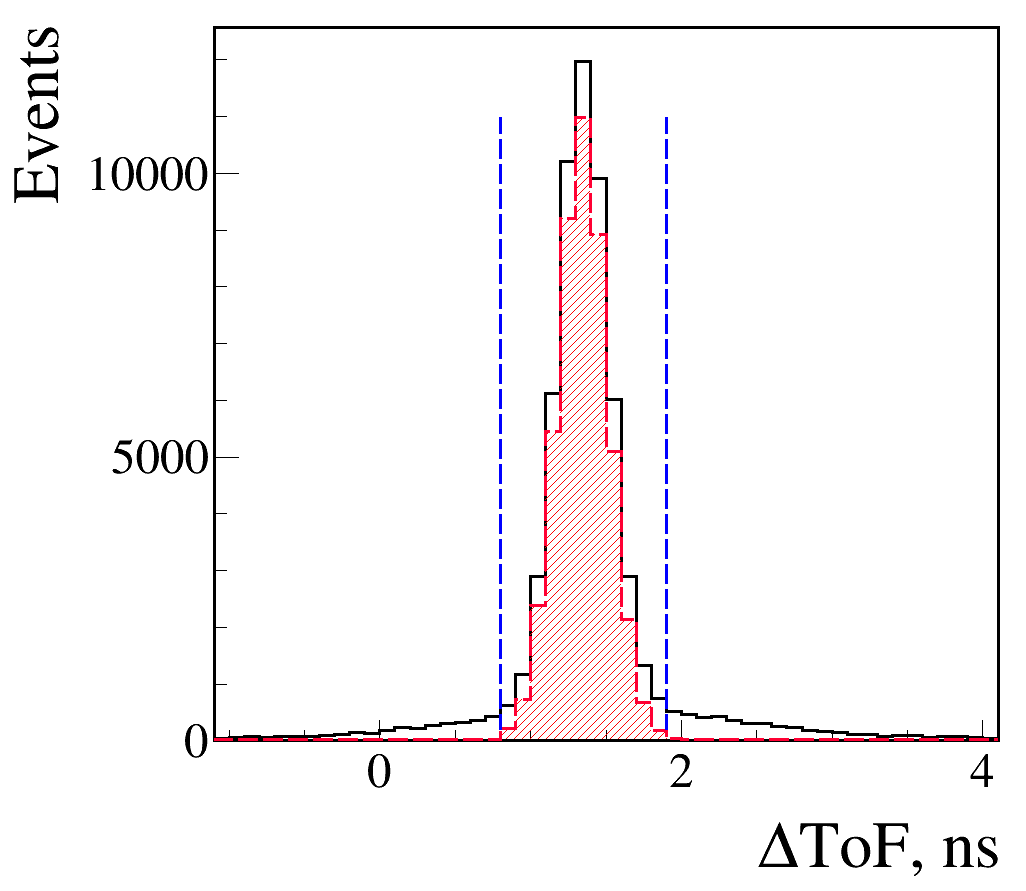}
     \caption{Time difference between the signals for the scattered and recoil proton detectors
      at a scattering angle of  75$^\circ$ in the c.m. and at 550 MeV/nucleon. 
     }
\label{fig:tdc_cut}
\end{figure}

The further selection of the events used to deduce the spin-dependent asymmetry was  performed using 
  the time difference between the signals for the conjugated proton detectors. 
Fig.~\ref{fig:tdc_cut}
shows the time difference between the signals for the scattered and recoil 
proton detectors.   The open histogram  represents the time difference  without the criteria on the
signal amplitude correlation.
The dashed histogram demonstrates the
events from the area inside the graphical cut shown in
Fig.~\ref{fig:adc_cut} by the solid line. 
The vertical dashed lines indicate the prompt timing window for the $pp$  quasi-elastic event selection. 
Only events within this window were included in the subsequent analysis.

\begin{figure}[hbtp]
    \centering
    \includegraphics[width=120mm]{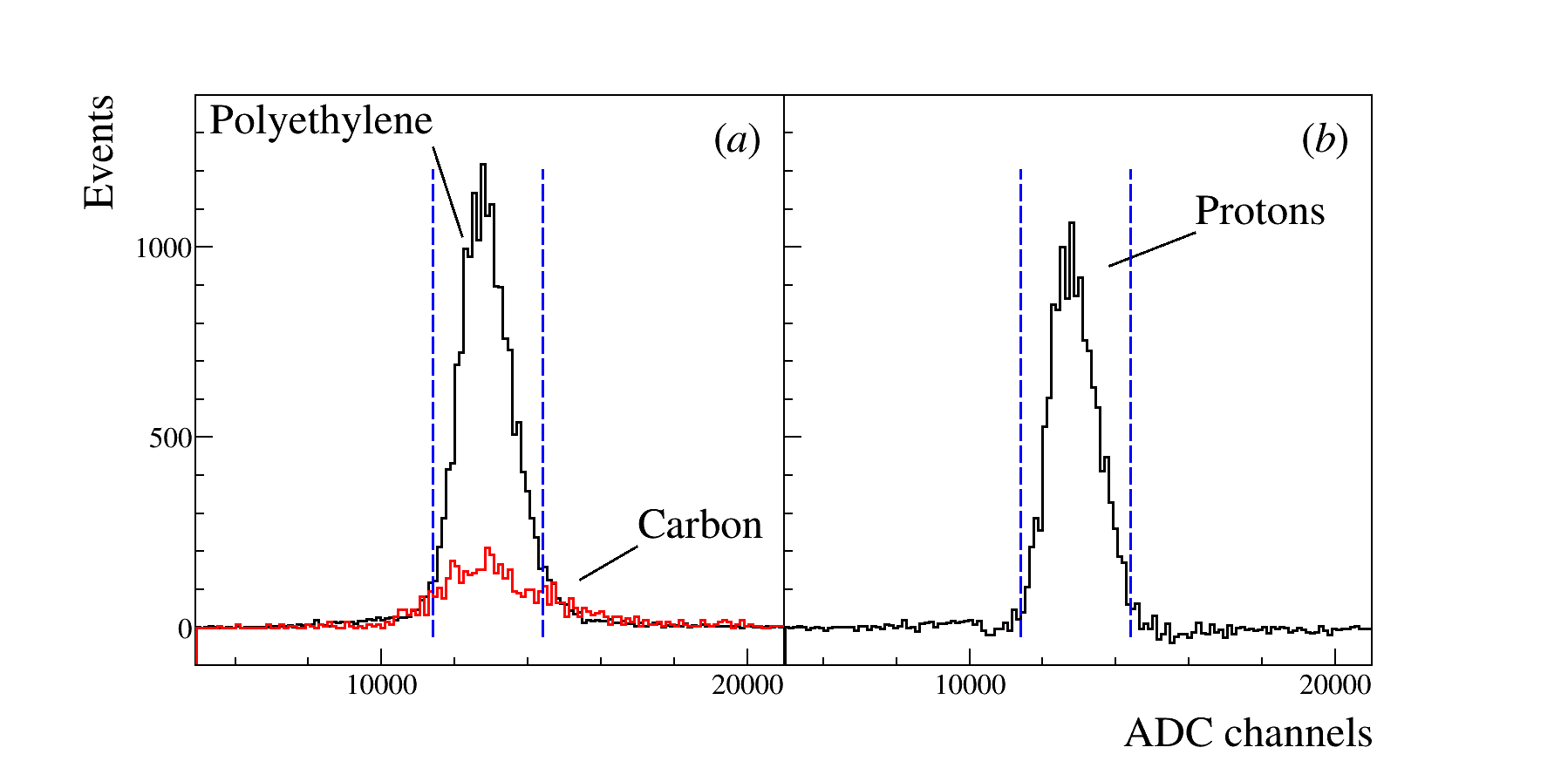}
     \caption{
      The $\text{CH}_2-\text{C}$  subtraction procedure at a scattering angle of  75$^\circ$ in the c.m. 
and at 550 MeV/nucleon.  
    }
\label{fig:subtr_cuts}
\end{figure}

The $\text{CH}_2-\text{C}$  subtraction procedure at  a scattering angle of  75$^\circ$ in the c.m. 
and at 550 MeV/nucleon is demonstrated in Fig.~\ref{fig:subtr_cuts}. 
The left panel shows normalized energy loss spectra for $\text{CH}_2$ and carbon targets.
The result of the carbon background subtraction  
is given in the right panel of Fig.~\ref{fig:subtr_cuts}. The vertical  dashed lines indicate the prompt 
windows for the final selection of the events obtained at the hydrogen content of $\text{CH}_2$.
Note that the vertical lines in Fig.~\ref{fig:subtr_cuts}  correspond to the dashed lines in Fig.~\ref{fig:adc_cut}.
A similar selection procedure was applied for the data obtained using the polarized proton beam at 500 MeV. 
The details of the $pp$ quasi-elastic event selection can be found in Refs~\cite{Volkov_2024_PPNL_Ay500, Volkov_2024_PhysAtNucl_Ay}.

\section{Measurement of the deuteron beam vector polarization}
\label{sec:dpol}

The  DSS experiment  ~\cite{dss1,dss2,dss3} was performed in two stages separated by a substantial  
time break  of about two weeks,  which was used for data taking  in the ALPOM-2 experiment ~\cite{basilev_2020}. 
The SPI settings for the spin modes "2--6" and "3--5" could have been altered over such a long time interval,
so we treat the data sets from these two stages as independent. 
Accordingly, the measurements at beam energies of 500 and 650 MeV/nucleon  and at 200 and 550 MeV/nucleon belong to the first and to the second parts of the experiment, respectively.  
The vector component of the deuteron beam polarization at these energies 
was determined  using the asymmetry of the
$pp$-quasi-elastic scattering yields and the known analyzing power $A_y$.   
Since the analyzing powers of elastic and quasi-elastic $pp$- scattering show no difference within the experimental precision achieved over a broad energy range ~\cite{Ball_1999_pp_el_quasiel}, the values of the analyzing power $A_y$  used
to evaluate the beam polarization were taken equal to those for $pp$-  elastic scattering.
The values of the analyzing power $A_y$  at these energies as a function of the scattering angle in the c.m. 
were approximated by using the following function: 
\begin{eqnarray}
    A_y (\theta) = a_1 \left(\theta - \frac{\pi}{2}\right) + a_2 \left(\theta - \frac{\pi}{2}\right)^3
    \label{eq:ay_fit}
\end{eqnarray}
where $\theta$ is the scattering angle in the c.m. taken in radians, and
$a_1$ and $a_2$ are the parameters of the approximation.
This function was chosen to ensure $A_y (\frac{\pi}{2}) = 0$ and to preserve the anti-symmetric behavior 
of $A_y$ around $\frac{\pi}{2}$, which follows from the symmetry properties of $pp$- elastic scattering.

\begin{figure}[hbtp]
    \centering
    \includegraphics[width=120mm]{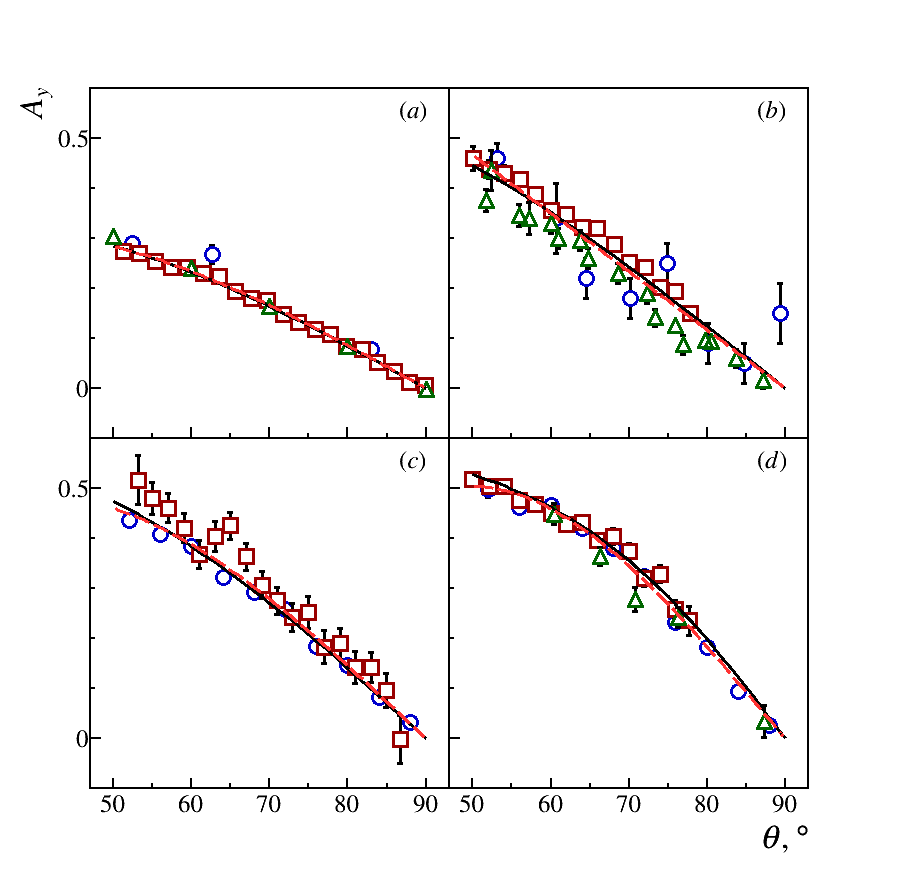}
     \caption{ Analyzing power of $pp$- elastic scattering at  $\sim$200 MeV,  $\sim$500 MeV,  $\sim$550 MeV, and $\sim$650 MeV is shown in  panels (\textit{a}), (\textit{b}), (\textit{c}), and (\textit{d}), respectively.
The symbols represent the  world experimental data; the solid and dashed lines are explained in the text. 
    }
    \label{fig:Ay_fits}
\end{figure} 

The results of the approximation of the experimental data on the analyzing power $A_y$ of  $pp$-elastic scattering at  $\sim$200 MeV,  $\sim$500 MeV,  $\sim$550 MeV, and $\sim$650 MeV  are demonstrated in panels (\textit{a}), 
(\textit{b}), (\textit{c}), and (\textit{d}) of  Fig.\ref{fig:Ay_fits}. 
The circles, triangles, and squares in Fig.\ref{fig:Ay_fits}(\textit{a}) are the data obtained at the energies around 
200 MeV from Refs
\cite{Baskir1957}, \cite{Tinlot1961}, and \cite{Rathmann1998}, respectively. 
The circles, triangles, and squares in Fig.\ref{fig:Ay_fits}(\textit{b})
represent the data at $\sim$500 MeV from Refs 
\cite{Albrow1970},  \cite{Cozzika1967}, and \cite{Bystricky1985}, respectively. 
The circles and squares are the data obtained at $\sim$550 MeV  from
Refs \cite{Altmeier2005} and \cite{Ball1987}, respectively. 
The circles, squares, and triangles represent the data at $\sim$650 MeV from Refs
\cite{Altmeier2005}, \cite{Bystricky1985}, and \cite{Glass1993}, respectively. The
solid lines are the results of the solution SP07 of the SAID partial  wave analysis (PWA) \cite{said2007}. 
The dashed lines  represent the  parametrization using function~\eqref{eq:ay_fit}. 
One can see good description of the world data on the analyzing power of $pp$-  elastic scattering by the 
function~\eqref{eq:ay_fit}, as well as good agreement of the parametrization and PWA SP07 solution \cite{said2007}. 
The results on the $a_1$ and $a_2$  parameters of the  function~\eqref{eq:ay_fit} for the  200, 500, 550, and 650 MeV are listed in Table \ref{tab:Ay_fit_params}. 

\begin{table}[hbtp]
    \centering
  \caption{
        Parameter values of the  function~\eqref{eq:ay_fit} for $pp$- elastic scattering at the beam energies of 200, 500, 550, and 650 MeV/nucleon.
    }
    \begin{tabular}{cccc}
        \hline
        Energy,      & $a_1\pm\Delta a_1$,     & $a_2\pm\Delta a_2$,   & $\chi^2$/NDF \\
        MeV/nucleon  & $\text{rad}^{-1}$ &  $\text{rad}^{-3}$ &   \\        
       \hline
        200 & -0.497$\pm$0.006 & 0.188$\pm$0.017 & 2.040 \\
        500 & -0.669$\pm$0.014 & 0.003$\pm$0.048 & 3.612 \\ 
        550 & -0.844$\pm$0.020 & 0.379$\pm$0.071 & 1.618 \\
        650 & -1.076$\pm$0.016 & 0.725$\pm$0.049 & 1.352 \\
        \hline
    \end{tabular}
    \label{tab:Ay_fit_params}
\end{table}

The  $pp$-quasi-elastic scattering yields in the "2--6" and "3--5" spin modes are
normalized to the yield for the unpolarized mode after correction
for the integrated beam intensity and dead time of the DAQ
system. For this purpose, the absolute value of the beam intensity is
not necessary, and the relative value monitored by the $pp$-quasi-elastic  
scattering at 90$^\circ$ in the c.m. is sufficient.
The normalized yields of the $pp$-quasi-elastic events for the left (L) 
and right (R) scattering~\cite{Ohlsen1972}:
\begin{eqnarray} 
    L &=& 1 + p_y\cdot A_y, \label{eq:N_L}  \\
    R &=& 1 - p_y\cdot A_y,
    \label{eq:N_R}
\end{eqnarray}
were used to determine the beam polarization.
Here, $p_y$ is the vector beam polarization  
and $A_y$ is the vector analyzing power of $pp$- quasi-elastic scattering.

\begin{figure}[hbtp]
    \centering
    \includegraphics[width=120mm]{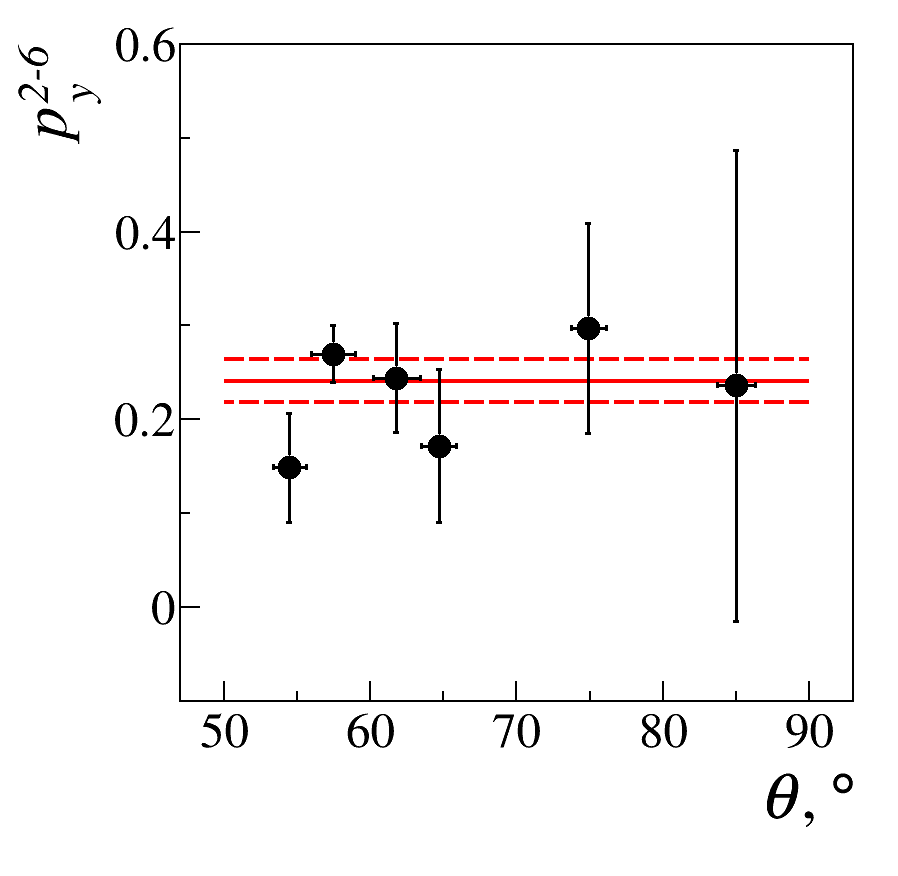}
     \caption{ Deuteron beam vector polarization  for the  spin mode "2--6" as a function
of the proton scattering angle at 650 MeV.
    }
    \label{fig:py26_angles}
\end{figure}

\begin{figure}[hbtp]
    \centering
    \includegraphics[width=120mm]{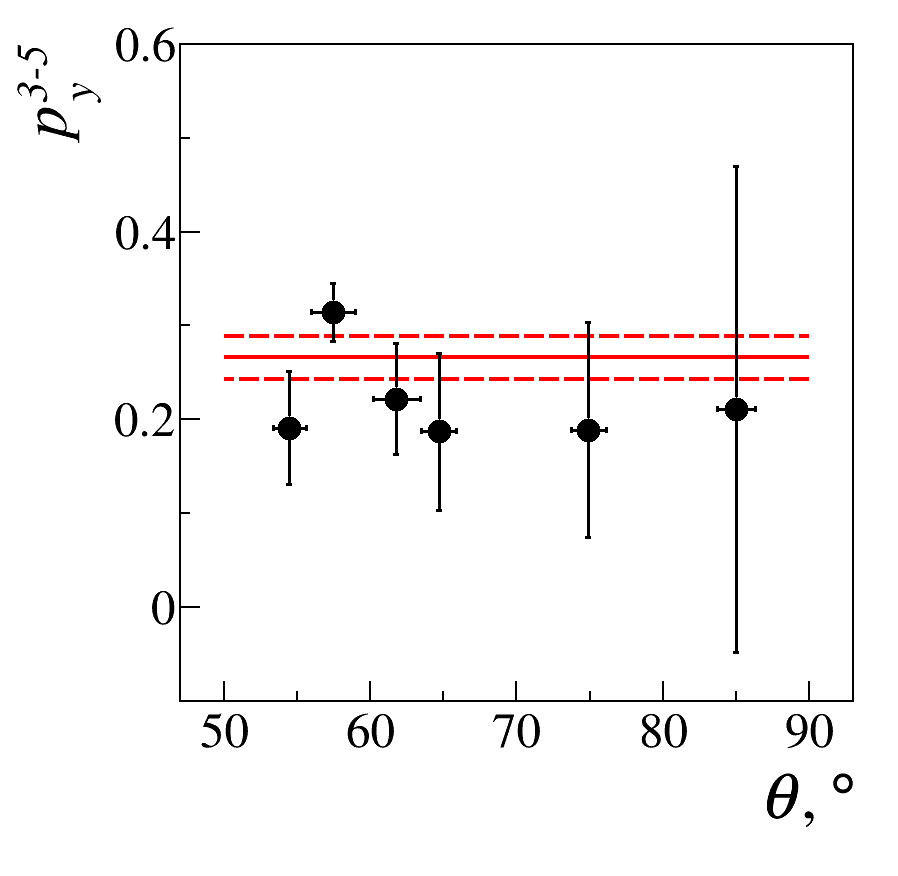}
     \caption{ Deuteron beam vector polarization  for the spin mode "3--5" as a function
of the proton scattering angle at 650 MeV.
    }
    \label{fig:py35_angles}
\end{figure}

The values of the deuteron beam vector polarization   for the SPI spin modes "2--6" and "3--5"  
were obtained  using expressions (\ref{eq:N_L}) and (\ref{eq:N_R}) 
for 6 different scattering angles in the c.m. at the energies of 200, 500,
550, and 650 MeV/nucleon. 
The values of the analyzing power $A_y$ were taken from the expression  ~\eqref{eq:ay_fit} with the
parameter values listed in Table \ref{tab:Ay_fit_params}. 
The scattering angle values were corrected for the interaction point position \cite{Volkov_2024_PhysAtNucl_Ay}.
Figs. \ref{fig:py26_angles}  and \ref{fig:py35_angles} demonstrate 
the values of  $p_y^{2-6}$ and $p_y^{3-5}$   at 650 MeV 
as a functions of the proton scattering angle in the c.m.  
The error bars include both statistical and systematic errors, 
which are related to the uncertainties of the background subtraction. 
One can see  good agreement of the polarization values obtained at different scattering angles in the c.m. 
Relatively  large error bars at an angle of $\sim$85$^\circ$ are explained by the small values of $A_y$. 
The solid lines are the deuteron beam polarization values
averaged over all the scattering angles. The dashed lines indicate the $\pm$1$\sigma$ deviations.
 
The polarimeter figure of merit was increased significantly using the data at the six c.m. scattering angles given in Table~\ref{tab:detectors_prop}.
The deuteron beam polarization values averaged over all the  angles   with  the statistical and systematic uncertainties obtained at 200, 500, 550, and 650 MeV/nucleon are  given in Table~\ref{tab:py_all_energies}.
Systematic uncertainties arise from the choice of the carbon background subtraction method.

\begin{table}[h]
    \caption{
        \label{tab:py_all_energies}
The values of the deuteron beam vector polarization with the statistical ($\Delta p_{y}^\text{stat}$) 
and systematic ($\Delta p_{y}^\text{sys}$) uncertainties at 200, 500, 550, and 650 MeV/nucleon. 
    }
    \centering
    \begin{tabular}{ccccc}
        \hline
             Beam energy, 
             &Spin mode
             &$p_{y}$
             &$\Delta p_{y}^\text{stat}$
             &$\Delta p_{y}^\text{sys}$\\ 
             MeV/nucleon & & & &\\
        \hline
            500 & "2--6" & 0.248 & 0.012 & 0.009 \\
                & "3--5" & 0.246 & 0.013 & 0.012 \\
        \hline
            650 & "2--6" & 0.241 & 0.023 & 0.008 \\
                & "3--5" & 0.266 & 0.023 & 0.008 \\
        \hline
            200 & "2--6" & 0.184 & 0.048 & 0.047 \\
                & "3--5" & 0.224 & 0.048 & 0.081 \\
        \hline
            550 & "2--6" & 0.214 & 0.018 & 0.016 \\
                & "3--5" & 0.271 & 0.017 & 0.015 \\
        \hline
    \end{tabular}
\end{table}

The deuteron beam polarization  was  measured several times using the polarimeter 
based on spin-asymmetry measurements   
in $dp$- elastic scattering at large
angles and at a deuteron kinetic energy of 270 MeV ~\cite{dp270}.  
The accuracy of  both vector and tensor components of the deuteron beam polarization  
achieved with this method is
better than 2\%. 
The averaged values of the  deuteron beam vector polarization 
for the   spin modes "2--6" and "3--5"  for the two parts of the DSS experiment
obtained using the deuteron beam polarimeter at 270 MeV ~\cite{dp270,Skhomenko2019} are given in Table~\ref{tab:polarization_dp_270mev}. 

\begin{table}[hbtp]
    \caption{
        \label{tab:polarization_dp_270mev}
        Deuteron beam vector polarization obtained using the deuteron beam polarimeter at 270 MeV ~\cite{dp270,Skhomenko2019} for the  spin modes "2--6" and "3--5".
    }
    \centering
   \begin{tabular}{cccc}
        \hline
            Experiment& Spin mode&$p_y$&$\Delta p_y$\\
            part      &     &    & \\
        \hline
            1& "2--6" & 0.231 & 0.008\\
             & "3--5" & 0.245 & 0.006\\
        \hline
            2& "2--6" & 0.212 & 0.007\\
             & "3--5" & 0.239 & 0.005\\
        \hline
    \end{tabular}
\end{table}

\begin{figure}[hbtp]
    \centering
    \includegraphics[width=120mm]{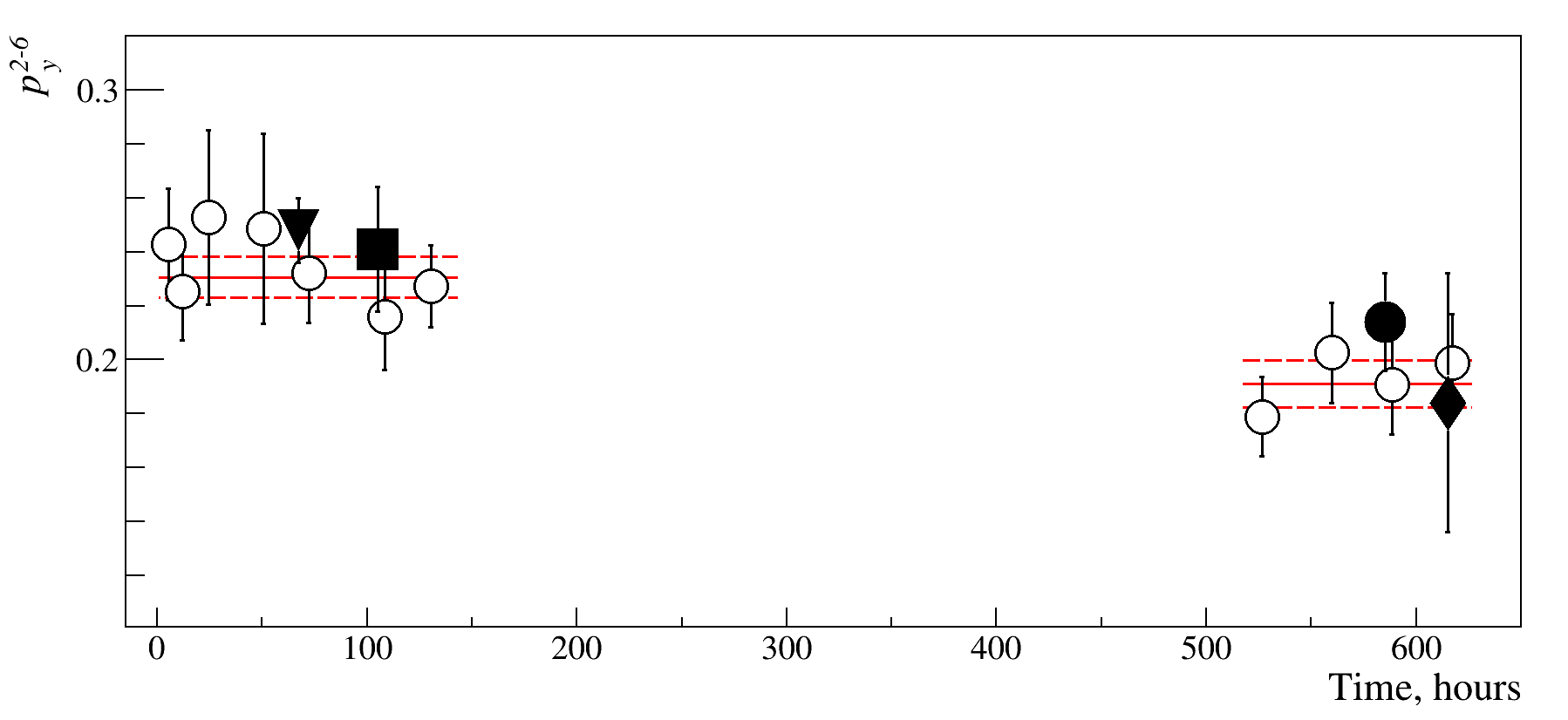}
     \caption{Vector polarization of the deuteron beam for the spin mode "2--6"
versus the measurement time in hours. 
    }
    \label{fig:py26_time}
\end{figure} 

 \begin{figure}[hbtp]
    \centering
    \includegraphics[width=120mm]{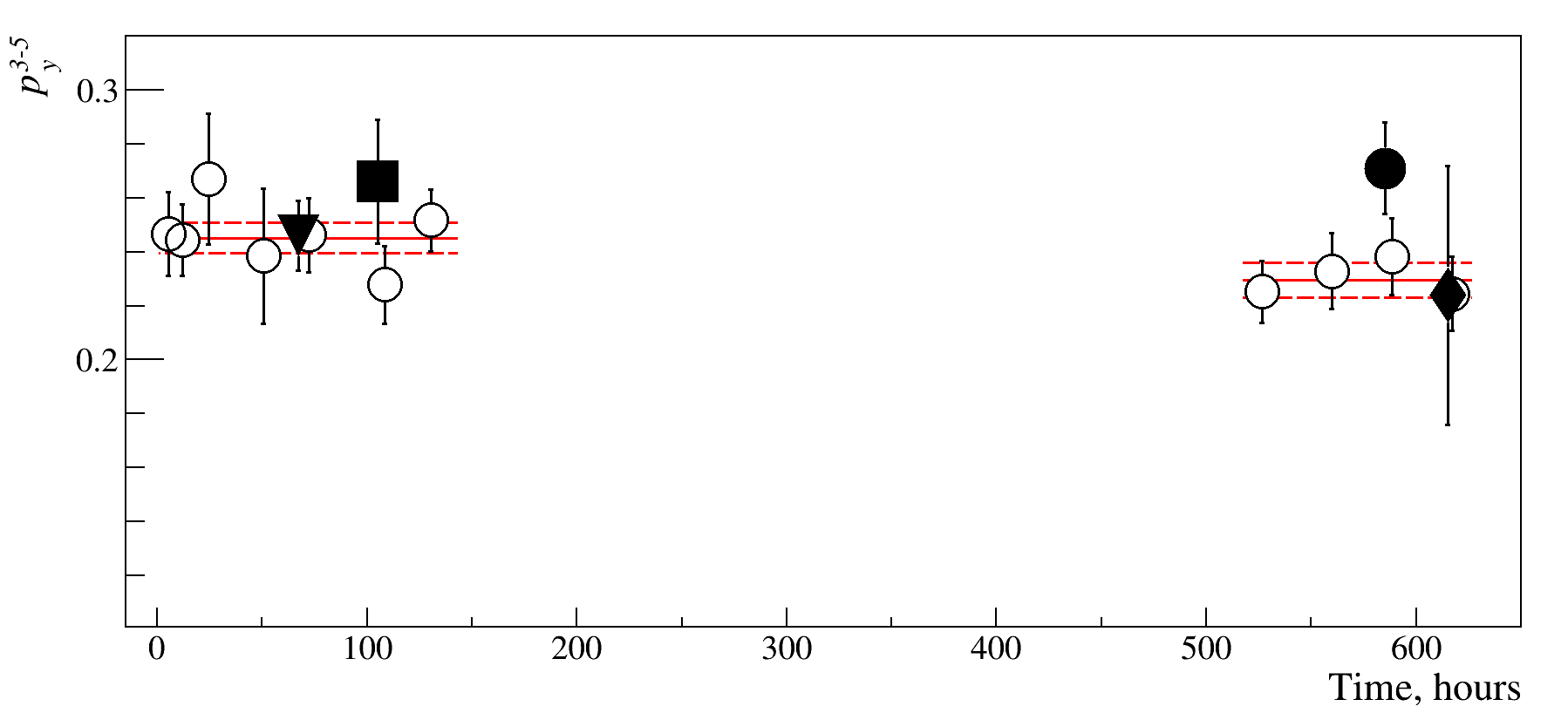}
     \caption{Vector polarization of the deuteron beam for the spin mode "3--5"
versus the measurement time in hours. 
    }
    \label{fig:py35_time}
\end{figure}

Vector polarizations of the  deuteron beam for the spin modes "2--6" and "3--5"
versus the measurement time in hours are presented in Figs.~\ref{fig:py26_time}
and \ref{fig:py35_time}, respectively. Open circles are the data obtained with the 
tensor-vector deuteron beam polarimeter at 270 MeV ~\cite{Skhomenko2019}.
The solid lines are the averaged vector polarization values from Table ~\ref{tab:polarization_dp_270mev}. 
Dashed lines show the $\pm$1$\sigma$ of the statistical errors. 
The solid triangles, squares, circles, and diamonds correspond to the results obtained using $pp$- 
quasi-elastic scattering at  different deuteron beam energies (in chronological order): 
500, 650, 550, and 200 MeV/nucleon, respectively.
One can see good agreement between the results obtained by two methods of measuring the vector polarization of the deuteron beam. 
  

\section{Effective analyzing power and figure of merit}
\label{sec:FoM}

The spin- dependent asymmetries were calculated using three different selection procedures for the events.
The first method used normalized yields of the $pp$- quasi-elastic events after 
CH$_2$--C subtraction. These asymmetries were used to evaluate the vector polarization of the deuteron beam described 
in Section~\ref{sec:dpol}.  The second method of the event selection also applied the cuts on the  
interaction point information \cite{TPM}, the energy loss correlations, and time difference for the kinematically
conjugated detectors, but without  CH$_2$--C subtraction.  
Therefore, in this case, the normalized yield  contains both $pp$- and $p\text{C}$-quasi-elastic events.
The third method  applied  was to consider the yield for the kinematically
conjugated detectors without cuts. 
This leads to the contribution of inelastic events as well.  

\begin{figure}[hbtp]
    \centering
    \includegraphics[width=120mm]{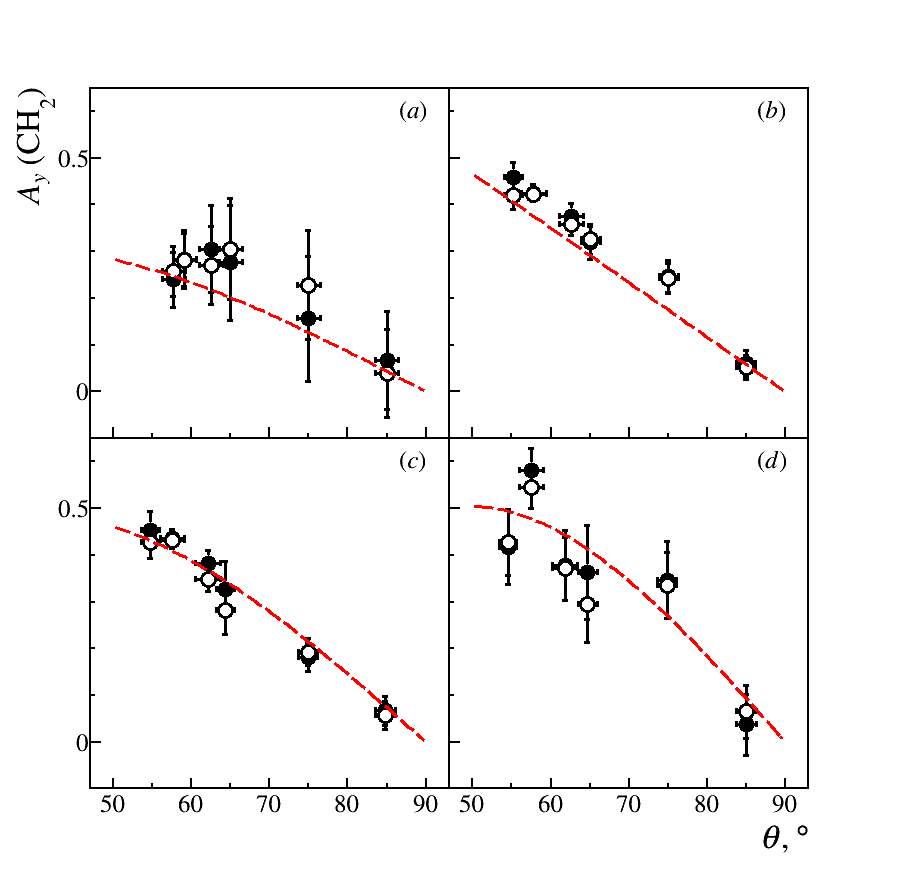}
     \caption{Analyzing power of $p\text{CH}_2$-scattering at  $\sim$200 MeV,  $\sim$500 MeV,  $\sim$550 MeV, and $\sim$650 MeV is shown in panels (\textit{a}), (\textit{b}), (\textit{c}), and (\textit{d}), respectively. The symbols and lines are explained in the text.
    }
    \label{fig:AyCH2}
\end{figure}

\begin{table}[hbtp]
    \centering  
     \caption{
        Analyzing power $A_y$  for the $p\text{CH}_2$ reaction with the use of the selection cuts.   
        The scattering  angle values in the c.m. are corrected  according to
        Ref.~\cite{Volkov_2024_PhysAtNucl_Ay}.
    }
    \begin{tabular}{ccccc}
        \hline
             Kinetic energy, 
             &$\theta$, $^\circ$
             &$\Delta \theta$,  $^\circ$
             &$A_{y}$
             &$\Delta A_{y}^{stat}$\\ 
             MeV/nucleon &   &   & & \\
        \hline
            200 & 57.7 & 1.6 & 0.240 & 0.059 \\
                & 59.1 & 1.3 & 0.284 & 0.062 \\
                & 62.5 & 1.6 & 0.305 & 0.093 \\
                & 65.0 & 1.2 & 0.276 & 0.123 \\
                & 75.0 & 1.3 & 0.156 & 0.133 \\
                & 85.0 & 1.3 & 0.067 & 0.104 \\
        \hline
            500 & 55.2 & 1.2 & 0.458 & 0.033 \\
                & 57.8 & 1.6 & 0.425 & 0.018 \\
                & 62.6 & 1.5 & 0.375 & 0.027 \\
                & 65.1 & 1.2 & 0.319 & 0.035 \\
                & 75.1 & 1.2 & 0.246 & 0.034 \\
                & 85.0 & 1.3 & 0.060 & 0.028 \\
        \hline
            550 & 54.7 & 1.1 & 0.454 & 0.039 \\
                & 57.5 & 1.5 & 0.436 & 0.019 \\
                & 62.1 & 1.6 & 0.382 & 0.028 \\
                & 64.3 & 1.2 & 0.327 & 0.059 \\
                & 74.9 & 1.2 & 0.181 & 0.030 \\
                & 84.8 & 1.3 & 0.067 & 0.031 \\
        \hline
            650 & 54.5 & 1.1 & 0.417 & 0.080 \\
                & 57.5 & 1.5 & 0.581 & 0.048 \\
                & 61.8 & 1.6 & 0.378 & 0.075 \\
                & 64.7 & 1.2 & 0.363 & 0.100 \\
                & 74.9 & 1.2 & 0.347 & 0.083 \\
                & 85.0 & 1.3 & 0.037 & 0.065 \\
        \hline
    \end{tabular}
    \label{tab:Ay_CH2with}
\end{table}
 
\begin{table}[hbtp]
    \centering  
     \caption{
        Analyzing power $A_y$  for the $p \text{CH}_2$ reaction without the selection cuts.   
        The scattering  angle values in the c.m. are corrected  according to
        Ref.~\cite{Volkov_2024_PhysAtNucl_Ay}.
    }
    \begin{tabular}{ccccc}
        \hline
             Kinetic energy, 
             &$\theta$, $^\circ$
             &$\Delta \theta$,   $^\circ$
             &$A_{y}$
             &$\Delta A_{y}^{stat}$\\ 
             MeV/nucleon &   &  & & \\
        \hline
            200    
& 57.7 & 1.6 &      0.258  &  0.053   \\
& 59.1 & 1.3 &      0.282  &  0.056   \\  
& 62.5 & 1.6 &      0.270  &  0.083   \\ 
& 65.0 & 1.2 &      0.305  &  0.108   \\ 
& 75.0 & 1.3 &      0.228  &  0.117   \\  
& 85.0 & 1.3 &      0.039  &  0.094   \\ 
        \hline
500  
& 55.2  &   1.2  &   0.420 &   0.030   \\             
& 57.8  &   1.6  &   0.422 &   0.017   \\             
& 62.6  &   1.5  &   0.359 &   0.025   \\             
& 65.1  &   1.2  &   0.326 &   0.032   \\             
& 75.1  &   1.2  &   0.243 &   0.032   \\             
& 85.0  &   1.3  &   0.052 &   0.026   \\  
         \hline
550  
& 54.7  &   1.1  &   0.428  &   0.035   \\      
& 57.5  &   1.5  &   0.432  &   0.018   \\      
& 62.1  &   1.6  &   0.349  &   0.026   \\       
& 64.3  &   1.2  &   0.281  &   0.050   \\       
& 74.9  &   1.2  &   0.193  &   0.029   \\       
& 84.8  &   1.3  &   0.058  &   0.030   \\   
         \hline   
650
& 54.5  &  1.1  &   0.427   &    0.070    \\        
& 57.5  &  1.5  &   0.546   &    0.045     \\        
& 61.8  &  1.6  &   0.371   &    0.067     \\        
& 64.7  &  1.2  &   0.295   &    0.082     \\        
& 74.9  &  1.2  &   0.336   &    0.071     \\        
& 85.0  &  1.3  &   0.065   &    0.057     \\      
       \hline
    \end{tabular}
    \label{tab:Ay_CH2without}
\end{table}

The values of the analyzing powers were 
obtained on the polyethylene $A_y(\text{CH}_2)$ at  200 MeV,  500 MeV,  550 MeV, and  650 MeV  calculated 
using expressions \eqref{eq:N_L} and \eqref{eq:N_R} with the values of the deuteron beam vector polarization 
obtained by the  tensor-vector  deuteron beam polarimeter at 270 MeV ~\cite{dp270,Skhomenko2019} 
(see Table~\ref{tab:polarization_dp_270mev}). 
The analyzing powers  $A_y(\text{CH}_2)$ at  200 MeV,  500 MeV,  550 MeV, and  650 MeV  
are demonstrated in panels (\textit{a}),  (\textit{b}), (\textit{c}), and (\textit{d}) of  Fig.\ref{fig:AyCH2}. 
The open and filled circles correspond to the data obtained on the CH$_2$ target with and without the use of the selection cuts. The lines are the results of the parametrization of the $pp$-elastic scattering 
analyzing power   using the function~\eqref{eq:ay_fit} with the
parameter values listed in Table \ref{tab:Ay_fit_params}. 
The data obtained on CH$_2$ with and without the use of the selection cuts 
are listed in Tables \ref{tab:Ay_CH2with} and \ref{tab:Ay_CH2without}, respectively. 
The good agreement of the data sets with different
selection cuts demonstrates that the contribution of inelastic channels is small at these energies. 
On the other hand, the $A_y(\text{CH}_2)$ data  are well described by the  parameterization of the 
$pp$-elastic scattering  analyzing power. 
Therefore, the data obtained on the  polyethylene target (without CH$_2$--C subtraction)
can be used for the polarimetry purposes in this energy range. 
Moreover, the scattering off the CH$_2$ target can be successfully used in the fast on-line mode,
e.g. without data storage and further off-line analysis.
The statistical errors achieved are large enough due to the short measurement time. 
Therefore,  the calibration of the polarimeter at different energies is required
in order to apply  the $p\text{CH}_2$- scattering   for the beam polarimetry.

The performance of the polarimeter is expressed in terms of
the figure of merit, ${\cal F}$. The counting rate
$N_{inc}$, needed for the desired beam polarization accuracy $\Delta P$, can be 
evaluated from the expression:
\begin{equation} 
    \Delta P \sim \frac{\sqrt{2}}{{\cal F}\cdot \sqrt{N_{inc}}}  
\label{eq:dP}
\end{equation}
The figure of merit ${\cal F}$ is defined as a function of efficiency $\epsilon$ and analyzing power $A$ 
\begin{equation} 
    {\cal F} =\int \epsilon \cdot A^2 \cdot d\Omega,
\label{eq:FoM}
\end{equation} 
where $d\Omega$ is the solid angle, and integration   is carried out
over the angular coverage of the polarimeter; efficiency $\epsilon$=$N_{ev}$/$N_{inc}$ ,
where $N_{ev}$ is the number of useful events detected and $N_{inc}$ is the
number of incident particles, and $A$ is the analyzing power.
The number of incident particles $N_{inc}$ can be
evaluated from the number of $pp$-quasi-elastic events at 90$^\circ$ in the
c.m. detected within the solid angle of 9.57$\cdot$10$^{-3}$ sr. 
Estimations of the averaged luminosity ${\cal L}$ and of the luminosity integrated over   time  ${\cal L}^{int}$ 
during the measurements of the deuteron beam vector polarization
at four energies are presented in 
Table \ref{tab:Lum_d}. 

\begin{table}[h]
    \caption{
        \label{tab:Lum_d}
         Estimations of the averaged   ${\cal L}$  and of the luminosity integrated over time ${\cal L}^{int}$
         during the deuteron beam vector polarization measurement.
    }
    \centering
    \begin{tabular}{c c c}
        \hline
        Energy, & ${\cal L}$, & ${\cal L}^{int}$ \\
        MeV     & cm$^{-2}$s$^{-1}$ &  cm$^{-2}$ \\
        \hline
        200 & 2.20$\cdot$10$^{29}$  & 6.4$\cdot$10$^{32}$   \\
        500 & 1.12$\cdot$10$^{30}$  & 4.3$\cdot$10$^{33}$     \\
        550 & 1.44$\cdot$10$^{30}$  & 4.8$\cdot$10$^{33}$   \\
        650 & 2.64$\cdot$10$^{29}$  & 7.0$\cdot$10$^{32}$ \\
        \hline
    \end{tabular}
\end{table}

\begin{table}[hbtp]
    \caption{
        \label{tab:FoM1}
        Figure of merit of the Nuclotron internal target polarimeter for $pp$-quasi-elastic scattering.
    }
    \centering
    \begin{tabular}{c c c}
        \hline
        Energy & ${\cal F}$ & $\Delta {\cal F}$ \\
        \hline
        200 & 0.0144 & 0.0029 \\
        500 & 0.0270 & 0.0012 \\
        550 & 0.0286 & 0.0014 \\
        650 & 0.0329 & 0.0027 \\
        \hline
    \end{tabular}
\end{table}

\begin{table}[hbtp]
    \caption{
        \label{tab:FoM2}
        Figure of merit of the Nuclotron internal target polarimeter for $\text{pCH}_2$-quasi-elastic scattering (with the use of the selection cuts).      }
    \centering
    \begin{tabular}{c c c}
        \hline
        Energy & ${\cal F}$ & $\Delta {\cal F}$ \\
        \hline
        200 & 0.0187 & 0.0025 \\
        500 & 0.0317 & 0.0010 \\
        550 & 0.0292 & 0.0010 \\
        650 & 0.0364 & 0.0024 \\
        \hline
    \end{tabular}
\end{table}

\begin{table}[hbtp]
    \caption{
        \label{tab:FoM3}
        Figure of merit of the Nuclotron internal target polarimeter for $\text{pCH}_2$-interaction  (without the use of the selection cuts).       }
    \centering
    \begin{tabular}{c c c} 
       \hline
        Energy & ${\cal F}$ & $\Delta {\cal F}$ \\
        \hline
        200 & 0.0217 & 0.0026 \\
        500 & 0.0330 & 0.0010 \\
        550 & 0.0302 & 0.0010 \\
        650 & 0.0384 & 0.0025 \\
        \hline
    \end{tabular}
\end{table}

\begin{figure}[hbtp]
    \centering
    \includegraphics[width=120mm]{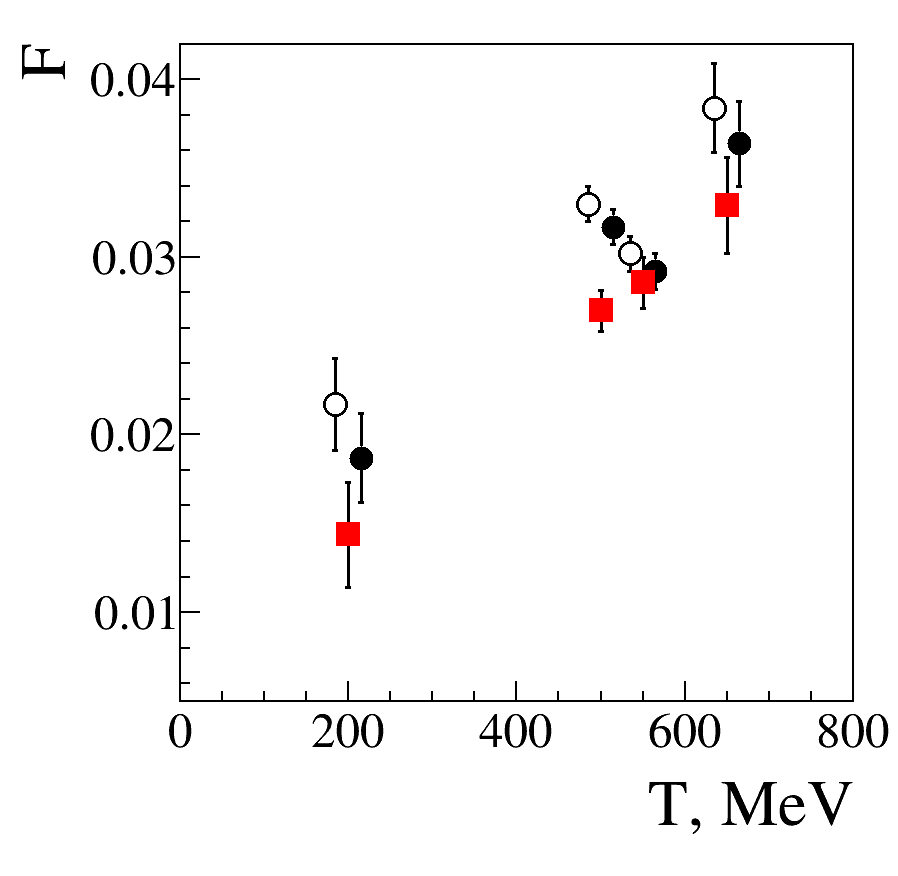}
     \caption{Figure of merit  of $pp$-quasi-elastic scattering and
$p\text{CH}_2$ interaction with and without the use of the selection cuts  as a function of incident energy.
    }
    \label{fig:FOM}
\end{figure} 

The estimation of the figures of merit ${\cal F}$ was performed as a sum over all 
detector pairs placed for the $pp$-quasi-elastic
scattering angles of $\sim$55$^\circ$-85$^\circ$ in the c.m.
The figure of merit results for   $pp$-quasi-elastic scattering and for 
$\text{pCH}_2$ interaction with and without the use of the selection cuts are presented in Tables
\ref{tab:FoM1}, \ref{tab:FoM2} and \ref{tab:FoM3}, as well as in Fig.~\ref{fig:FOM} by the filled
squares, and filled and open circles, respectively. 
The figures of merit for three types of the event selection are almost the same due to the small 
carbon background contribution. ${\cal F}$ values increase with the energy due to  the 
corresponding analyzing power values rising in this energy domain.

\section{Measurements of the proton beam polarization} 
\label{sec:proton_beam_results}

SPI \cite{Fimushkin_2016_SPI,Belov_2017_SPI} can  also provide polarized protons.  
The polarized proton beam was accelerated by RFQ and LINAC LU20 up to
5 MeV and injected into the Nuclotron ring  for the first time \cite{proton_acc2017}.
The  protons  were then accelerated up to 500 MeV. 
SPI provided proton beam polarization using weak-field transition 1$\rightarrow$3 with the maximal theoretical  value of the polarization +1. 
Two SPI spin modes, unpolarized and "1--3", were changed cyclically and spill-by-spill in the 
following manner: one cycle with the unpolarized mode  and two subsequent cycles with the "1--3" mode. 
The typical beam intensity was 2-3$\cdot$10$^7$ and $\sim$1.5$\cdot$10$^8$ ppp for the polarized and
unpolarized cases, respectively.

The proton beam polarization values were obtained using the same method as during measurements of 
the deuteron beam  vector polarization  described in Section~\ref{sec:dpol}. 
The $pp$-quasi-elastic events for each spin mode were obtained using the interaction point information,
cuts on the correlation of the energy losses, time difference, and carbon content subtraction. 
The normalized yields of the $pp$-quasi-elastic events for the left  
and right  scattering (see Eqs.~(\ref{eq:N_L}) and (\ref{eq:N_R}) )
for the SPI spin mode "1--3" were used to evaluate 
the proton  beam polarization  for 6 different scattering angles in the c.m. at the energy of  500 MeV. 
The values of the analyzing power $A_y$ were also taken from the expression  ~\eqref{eq:ay_fit} with the
parameter values listed in Table \ref{tab:Ay_fit_params}.

\begin{figure}[h]
    \centering
    \includegraphics[width=120mm]{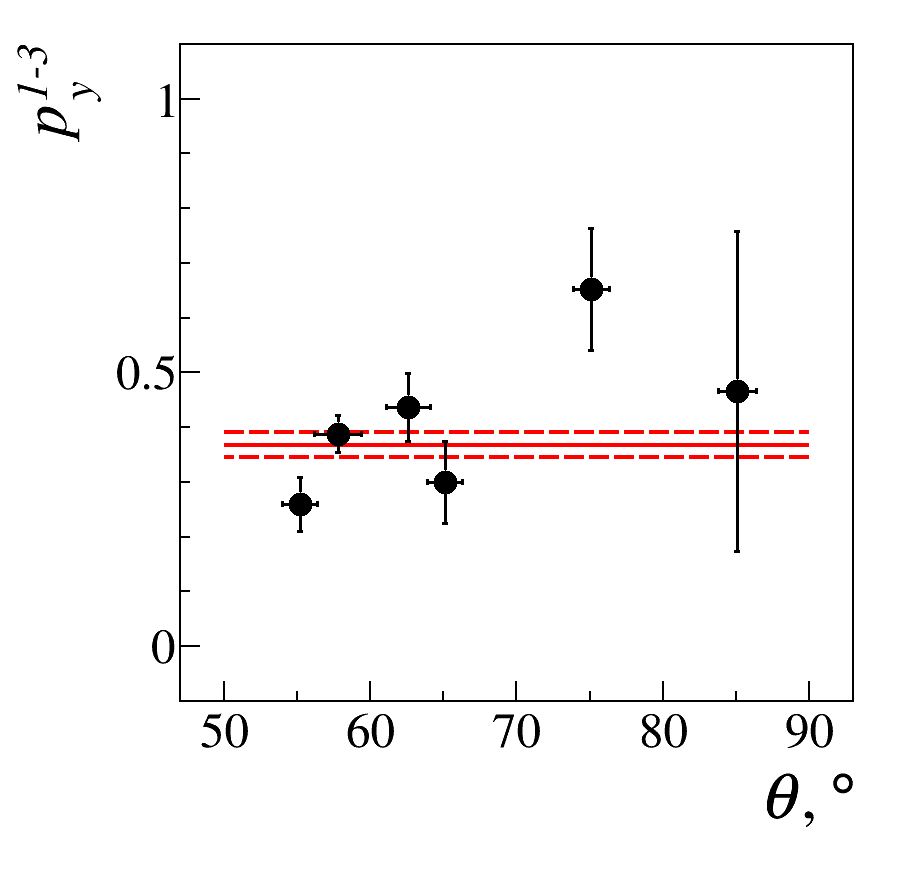}
    \caption{
        \label{fig:pol_500mev_protons}
 Proton beam polarization  for the  spin mode "1--3" as a function
of the proton scattering angle at 500 MeV. 
    }
\end{figure}

Proton beam polarization  for the spin mode "1--3" as a function
of the proton scattering angle at 500 MeV is shown in Fig. ~\ref{fig:pol_500mev_protons}.
The solid and dashed lines are the weighted average value and $\pm$1$\sigma$ of the standard deviation,
respectively. 
The  averaged value of the proton beam polarization was found as   $0.368 \pm 0.023$. 
The result for the false asymmetry (polarization)
for the unpolarized proton beam was found to be  consistent with zero: $0.038 \pm 0.023$. 
Further optimization of   SPI is required to increase the proton beam polarization value and intensity.
Another reason for the relatively small value of the proton beam polarization 
could be a possible depolarization effect at the integer resonance $\gamma G$=2 
at 108 MeV. Therefore,  systematic studies of the proton beam spin dynamics during transportation, injection and acceleration are necessary.

\section{Conclusion} 
\label{sec:conclusion}

A polarimeter based on $pp$- quasi-elastic scattering has been
installed at the internal target of the  Nuclotron. Measurements of
the deuteron beam vector polarization have been performed at several 
energies being in good agreement with the results obtained  
using the polarimeter based on  $dp$- elastic scattering  at 270 MeV \cite{dp270,Skhomenko2019}. 

The results on the effective analyzing powers of the proton scattering off the polyethylene target at
200, 500, 550, and 650 MeV/nucleon are presented. The angular behavior of the 
effective analyzing powers is consistent with the behavior of the $pp$- elastic scattering 
analyzing power and practically does not depend on the selection cuts.
Therefore, the proton scattering off the CH$_2$ target can be successfully used 
for the beam polarimetry, including the fast on-line mode in the energy range of 200--650 MeV/nucleon.

The polarimeter figure of merit ${\cal F}$ increases with  energy, approaching $\sim$0.03$\div$0.035 at  
650 MeV/nucleon. This value is significantly larger than the ${\cal F}_y$ value reported for the
tensor-vector polarimeter  based on  $dp$- elastic scattering  at 270 MeV \cite{dp270}.
Further improvement of the performance of the Nuclotron internal target  polarimeter  
by expanding the angular coverage and detector granularity  is planned to enhance the accuracy of 
polarization measurements 
\cite{Terekhin_2023_PEPAN_new_pol}.

The polarization of the polarized proton beam, accelerated for the first time  up to 500 MeV at 
the Nuclotron,   
was measured to be $\sim$0.37, while the value obtained with the unpolarized beam  
was close to zero. The relatively small value of the proton polarization can be related to both   
non-optimal settings of SPI and possible depolarizing influence of the  integer resonance $\gamma G$=2 
at 108 MeV. Further studies on   proton spin manipulation during  beam acceleration at the Nuclotron
are required \cite{Spin2025}.    
\\

{\bf Acknowledgments}\\

The authors are grateful to the Nuclotron staff  and the SPI group for providing good conditions for the experiment. They thank A.V. Butenko, A.D. Kovalenko, and A.O. Sidorin for providing 
the polarized proton beam, and A.S. Belov, V.B. Shutov and V.V Fimushkin for   tuning  SPI. 
They express their gratitude to S.N. Bazylev, A.N. Khrenov, V.I. Maximenkova, I.V. Slepnev,
V.M. Slepnev and A.V. Shutov for their help during the preparation of the detector and DAQ
system. The authors are grateful to Yu.N.Filatov and A.M.Kondratenko for   discussions
on   proton spin manipulation during injection into Nuclotron.

\bibliographystyle{elsarticle-num}


\begin{thebibliography}{99}

\bibitem{dss1} 
M.~Janek,  et al, Study of the dp Elastic and dp Breakup Complementary Processes Using Polarized and Unpolarized Beam of  Nuclotron, Few~Body~Syst. 63 (2022) 3,  https://doi.org/10.1007/s00601-021-01713-1.
   
\bibitem{dss2} 
V.P.~Ladygin,  et al., Angular Dependences of the Deuteron Analyzing Powers in Elastic dp Scattering at Large Transverse Momenta, Phys.Part.Nucl. 53 (2022) 251-255, https://doi.org/10.1134/S1063779622020459. 

\bibitem{dss3}   
P.K.~Kurilkin,   et al., Measurement of the vector and tensor analyzing powers for dp- elastic scattering at 880 MeV, Phys.Lett. B715 (2012) 61-65,   https://doi.org/10.1016/j.physletb.2012.07.022. 

\bibitem{basilev_2020} S.N.~Basilev,   et al., Measurement of neutron and proton analyzing powers on C, CH, CH$_2$ and Cu targets in the momentum region 3--4.2 GeV/c, Eur.Phys.J. A56 (2020) 26, 
https://doi.org/10.1140/epja/s10050-020-00032-z.


\bibitem{SPD_TDR} V.~Abazov, et al., 
{Technical Design Report of the Spin Physics Detector at NICA},  
Natural~Science~Review 1  (2024) 1,  
http://nsr-jinr.ru/index.php/nsr/article/view/35,  
e-Print: 2404.08317 [hep-ex].

\bibitem{SPD2025} V.P.~Ladygin,  Spin Physics Detector at NICA, Phys.Part.Nucl. 57
(2026) 423-427,  https://doi.org/10.1134/S1063779625701503.

\bibitem{nica}
V.D.~Kekelidze, et al., 
Three stages of the NICA accelerator complex,
Eur.Phys.J. A52 (2016) 211, 
http://doi.org/10.1140/epja/i2016-16211-2.

\bibitem{spd_gluon}
A.~Arbuzov,  et al., On the physics potential to study the gluon content of proton and deuteron at NICA SPD, Prog.Part.Nucl.Phys. 119 (2021) 103858, https://doi.org/10.1016/j.ppnp.2021.103858.
\bibitem{spd_stage1}
V.V.~Abramov,    et al., Possible Studies at the First Stage of the NICA Collider Operation with Polarized and Unpolarized Proton and Deuteron Beams,  Phys.Part.Nucl. 52 (2021) 6, 1044-1119,  
https://doi.org/10.1134/S1063779621060022.

\bibitem{Ball1987} J.~Ball, et al., 
Measurement of $np$ analyzing power $A_{oono}$ using the deuteron polarized beam of {Saturne II},
Nucl.Phys. B286 (1987) 635-642, https://doi.org/10.1016/0550-3213(87)90455-X.

\bibitem{Ball_1999_pp_el_quasiel} J.~Ball, et al., Elastic and quasi-elastic pp scattering in $^6$LiH and $^6$LiD targets between 1.1 and 2.4 {GeV},
Eur.Phys.J. C11 (1999) 51-67, https://doi.org/10.1007/s100529900149.

 
\bibitem{spinka} H.~Spinka, et al., Beam polarization at the ZGS,
Nucl. Instrum. Methods 11 (1983) 239-261,
https://doi.org/10.1016/0167-5087(83)90246-6.


\bibitem{kek} C.~Ohmori, et al., Proton polarimeter for high-energy experiments at KEK,
 Nucl. Instrum. Methods A 278 (1989) 705-712, 
https://doi.org/10.1016/0168-9002(89)91193-5.


\bibitem{saturn-2} A.N.~Prokofiev,  et al, Polarimetry of the polarized proton and deuteron beams at intermediate energies, Czech.J.Phys. 49S2 (1999) 29-36, 
https://doi.org/10.1007/s10582-999-0082-8.  


\bibitem{Altmeier2005} M.~Altmeier, et al., Excitation functions of the analyzing power in elastic proton-proton scattering from 0.45 to 2.5 {GeV}, Eur. Phys. J. A 23 (2005) 351-364, 
https://doi.org/10.1140/epja/i2006-10025-9.

\bibitem{azhgirey2003} L.S.~Azhgirey, et al., Intermediate-energy polarimeter for the measurement of the deuteron and proton beam polarization at the JINR Synchrophasotron, Nucl. Instrum. Methods  A 497 (2003) 340-349,   https://doi.org/10.1016/S0168-9002(02)01793-X. 

\bibitem{azhgirey2005}
L.S.~Azhgirey, et al., Measurement of the extracted deuteron beam vector polarization at the Nuclotron, Phys. Part.Nucl.Lett. 2 (2005) 122-127, https://doi.org/10.48550/arXiv.nucl-ex/0404023 

 
\bibitem{ITS} A.I.~Malakhov, et al., Potentialities of the internal target station at the Nuclotron, Nucl. Instrum. Methods   A 440  (2000) 320-329, https://doi.org/10.1016/S0168-9002(99)00966-3.

\bibitem{Fimushkin_2016_SPI} V.V.~Fimushkin, et al., Development of polarized ion source for the JINR accelerator complex, J.Phys.:Conf.Ser. 678 (2016) 012058,  https://doi.org/10.1088/1742-6596/678/1/012058.

\bibitem{Belov_2017_SPI} A.S.~Belov, et al.,
Source of polarized ions for the JINR accelerator complex, J.Phys.:Conf.Ser. 938 (2017) 012017,  
https://doi.org/10.1088/1742-6596/938/1/012017.

\bibitem{madison} The "Madison Convention", in Proceedings of the 3-rd International Symposium on Polarization Phenomena in Nuclear Reactions, Madison 1970, ed. by H.H. Barschall and W. Haeberli (University of Wisconsin Press, 1971), page xxv.

\bibitem{dp270} P.K.~Kurilkin,  et al., The 270 MeV deuteron beam polarimeter at the Nuclotron Internal Target Station, Nucl. Instrum. Methods  A 642  (2011) 45-51, 
https://doi.org/10.1016/j.nima.2011.03.054. 

\bibitem{ITS_DAQ} A.Yu.~Isupov, V.A.~Krasnov, V.P.~Ladygin, S.M.~Piyadin, and S.G.~Reznikov, The Nuclotron internal target control and data acquisition system, Nucl. Instrum. Methods   A 698, 127 (2013) 127-134,  https://doi.org/10.1016/j.nima.2012.09.057.
 
\bibitem{TPM} Yu.V.~Gurchin, et al., Target position monitor for the internal target station at the {Nuclotron}, Phys.Part.Nucl.Lett. 4, 3 (2007) 263-267, https://doi.org/10.1134/S1547477107030107.


\bibitem{DSS_DAQ} A.Yu.~Isupov, Online polarimetry of the {Nuclotron} internal deuteron and proton beams, 
J.Phys.:Conf.Ser. 938  (2017) 012019, https://doi.org/10.1088/1742-6596/938/1/012019.

\bibitem{tqdc16} https://afi.jinr.ru/TQDC-16.
\bibitem{TTCM} https://afi.jinr.ru/TTCM.
\bibitem{FVME} https://afi.jinr.ru/FVME.

\bibitem{FVME2}https://afi.jinr.ru/FVME2.
\bibitem{FVME2TMWR} https://afi.jinr.ru/FVME2TM.
\bibitem{U40VE-TM} https://afi.jinr.ru/U40VE.

\bibitem{Volkov_2024_PPNL_Ay500} I.S.~Volkov, et al., 
Vector Analyzing Power in Quasi-Elastic Proton-Proton Scattering at an Energy of 500 MeV/nucleon, 
Phys.Part.Nucl.Lett. 21, 1 (2024) 43-54, https://doi.org/10.1134/S1547477124010138.

\bibitem{Volkov_2024_PhysAtNucl_Ay}  I.S.~Volkov, et al., 
Analyzing Power Measurements of Quasi-Elastic Proton-Proton Scattering at the Intermediate Energies at the Nuclotron Internal Target, 
Phys.Atom.Nucl. 87, Suppl.3 (2024)  S438-S451,  https://doi.org/10.1134/S1063778824701072.

\bibitem{Baskir1957}  E.~Baskir,   E.M.~Hafner, A.~Roberts, and  J.H.~Tinlot,  
Polarization in Proton-Proton Scattering at 130, 170, and 210 {MeV},
Phys.Rev. 106 (1957) 564-568,  https://doi.org/10.1103/PhysRev.106.564.
\bibitem{Tinlot1961} J.H.~Tinlot, and R.E.~Warner,
Polarization in 217-{MeV} p-n and p-p Scattering,
Phys.Rev. 124 (1961)  890-896,  https://doi.org/10.1103/PhysRev.124.890.
\bibitem{Rathmann1998} F.~Rathmann, et al., Complete angular distribution measurements of $\mathrm{pp}$ spin correlation parameters ${A}_{\mathrm{xx}},$ ${A}_{\mathrm{yy}},$ and ${A}_{\mathrm{xz}}$ and analyzing power ${A}_{y}$ at 197.4 {MeV},
Phys.Rev. C58 (1998) 658-673, https://doi.org/10.1103/PhysRevC.58.658.

\bibitem{Albrow1970} M.G.~Albrow, et al., 
Polarization in elastic proton-proton scattering between 0.86 and 2.74 {GeV}/$c$,
Nucl.Phys. B23, 3 (1970) 445-465, https://doi.org/10.1016/0550-3213(70)90296-8.
\bibitem{Cozzika1967}  G.~Cozzika, et al., 
Measurements of the Polarization Parameters $P$ and  ${C}_{\mathrm{nn}}$ in 
$\mathrm{pp}$ Elastic Scattering between 500 and 1200 {MeV}, 
Phys.Rev. 164 (1967) 1672-1679, https://doi.org/10.1103/PhysRev.164.1672.
\bibitem{Bystricky1985} J.~Bystricky, et al.,
Measurement of the spin correlation parameter $\text{A}_\text{oonn}$ and of the analyzing power for $pp$ elastic scattering in the energy range from 0.5 to 0.8 GeV,
Nucl.Phys.  B262, 4 (1985) 727-743,  https://doi.org/10.1016/0550-3213(85)90513-9.



\bibitem{Glass1993}   G.~Glass, et al., 
Forward angle analyzing power in $\vec{p}-n$ and $\vec{p}-p$ quasifree scattering at 643 and 797 {MeV},
Phys.Rev. C47 (1993) 1369-1375,  https://doi.org/10.1103/PhysRevC.47.1369.

\bibitem{said2007} R.A.~Arndt,   W.J.~Briscoe,  I.I.~Strakovsky, and R.L.~Workman,
Updated analysis of {NN} elastic scattering to 3 {GeV},
Phys.Rev. C76 (2007) 025209, https://doi.org/10.1103/PhysRevC.76.025209.

\bibitem{Ohlsen1972}
 G.G.~Ohlsen,  Polarization transfer and spin correlation experiments in nuclear physics, Rept.Prog.Phys. 35 (1972) 717-801,  https://doi.org/10.1088/0034-4885/35/2/305.  

\bibitem{Skhomenko2019} Ya. T.~Skhomenko, et al., 
Deuteron beam polarimeter at Nuclotron internal target, EPJ Web Conf. 204 (2019) 10002, 
https://doi.org/10.1051/epjconf/201920410002.

\bibitem{proton_acc2017} S.V.~Barabin, et al., 
Injection of Polarized Protons and Light Ions in the Nuclotron Superconducting Synchrotron,
Phys.Part.Nucl.Lett. 15 (2018) 827-830, https://doi.org/10.1134/S1547477118070142.

\bibitem{Terekhin_2023_PEPAN_new_pol} A.A.~Terekhin, et al., 
Proton Polarimeter at the Internal Target Station of the Nuclotron at the Joint Institute for Nuclear Research, Phys. Part. Nucl. 54 (2023) 634-639, https://doi.org/10.1134/S1063779623040317.

\bibitem{Spin2025} V.P.~Ladygin, et al, Spin Physics Research INfractructure and
Technologies at NICA (SPRINT$@$NICA), PoS (SPIN2025) 152, https://pos.sissa.it/517/152/pdf.


\end{thebibliography}


\end{document}